# Number of natively unfolded proteins scales with genome size.

Antonio Deiana, Andrea Giansanti*

Physics Department, La Sapienza, University of Rome, P.le A. Moro, 2, 00185, Rome, Italy

* Correspondence to: Andrea.Giansanti@roma1.infn.it

## Abstract

Natively unfolded proteins exist as an ensemble of flexible conformations lacking a well defined tertiary structure along a large portion of their polypeptide chain. Despite the absence of a stable configuration, they are involved in important cellular processes. In this work we used from three indicators of folding status, derived from the analysis of mean packing and mean contact energy of a protein sequence as well as from VSL2, a disorder predictor, and we combined them into a consensus score to identify natively unfolded proteins in several genomes from Archaea, Bacteria and Eukarya. We found a high correlation among the number of predicted natively unfolded proteins and the number of proteins in the genomes. More specifically, the number of natively unfolded proteins scaled with the number of proteins in the genomes, with exponent $1.81 \pm 0.10$. This scaling law may be important to understand the relation between the number of natively unfolded proteins and their roles in cellular processes.

## Introduction

The existence of natively unfolded proteins is nowadays a well established experimental fact [1-5]. Natively unfolded proteins exist as an ensemble of flexible conformations, lacking a well defined tertiary structure along large portions of the polypeptide chain [4,6]. These proteins are involved in important cellular processes, like signalling, targeting and DNA binding [6-10]. It has been also suggested that they may play critical roles in cancer development [11] and in some amyloidotic diseases [12,13].
In this work we screened the genomes of several organisms from Archaea, Bacteria and Eukarya, searching for natively unfolded proteins. To identify these proteins, we combined into a consensus score three indicators of folding status, derived from the analysis of the mean packing [14,15] and mean contact energy [16] of the amino acid sequences, as well as from VSL2 [17,18], a disordered predictors that excellently performed at the recent experiment CASP7 [19]. In a previous work, we have shown that a consensus score is useful discriminate whether a protein is folded or unfolded by means of scalar indexes of fold; in particular we have introduced a strictly unanimous score able to resolve conflictual situations in which two folding indexes assign a protein to different folding classes [20]. We have shown that the strictly unanimous score had good performance in a test set of 743 folded and 81 natively unfolded proteins, selected from data reported in the literature, failing in classifying only about 10% of the proteins analysed. In this work we used the strictly unanimous score to search natively unfolded proteins in the genomes of several organisms. We found percentages of natively unfolded proteins consistent with those previously reported in the literature [21,22]. Moreover we observed a correlation between the number of predicted natively unfolded proteins and the number of proteins in the genomes. In logarithmic plot, the number of predicted natively unfolded proteins scaled with the number of proteins in the genomes, with exponent 1.81 ± 0.10. This scaling law, to be validated by further studies, may be important to understand the relation between the necessity to develop specific cellular processes and the number of natively unfolded proteins in the genomes.

## Methods

To predict the folding status of a protein, we used three indicators: mean packing, mean contact energy and an index derived from VSL2 that we called *gVSL2* [20].

The mean packing of a protein is the arithmetic mean of the packing values of its amino acids. The packing of an amino acid [14,15] is defined as the average number of its close residues, i.e. residues within a distance of 8 Å, computed on a large set of structured proteins. Natively unfolded proteins tend to have a lower mean packing than folded ones; in particular we considered natively unfolded amino acid sequences with a mean packing below 20.55.

The mean contact energy of a protein is the arithmetic mean of the contact energy values of its amino acids. The contact energy of an amino acid is a measure of its "contact interaction" with residues from 2 to 100 position apart, downward and upward, along the sequence. It is computed following the algorithm described by Dosztanyi *et al.* in [16]. Natively unfolded proteins tend to have a higher mean contact energy with respect to folded ones; we considered natively unfolded amino acid sequences with mean contact energy higher than -0.37 arbitrary energy unit (a.e.u.).

*gVSL2* is an index derived from disorder predictor VSL2 [17,18]; *gVSL2* is the arithmetic mean of the VSL2 scores, over the sequence. We considered a protein as natively unfolded if *gVSL2* was above 0.5.

The defined indexes are correlated, but we observed that, for several proteins, they disagreed in assigning a protein to a specific folding class. To resolve this conflictual situations, we introduced the strictly unanimous score $S_{SU}$ [20]. It required unanimous consensus among the indexes; more precisely, it classified a protein as folded if all the indexes predicted it as folded, conversely it classified a protein as natively unfolded if all the indexes predicted it as natively unfolded. If there was disagreement among at least two indexes, the strictly unanimous score left the proteins unclassified.

## Results

As said above, $S_{SU}$ requires consensus among mean packing, mean contact energy and *gVSL2* to assign a protein to a folding class. In a previous work we have shown that the strictly unanimous score has better performance than single folding indexes [20]. Moreover we have checked that the number of proteins left unclassified is generally low, about 10% of the analysed proteins; this suggests that the $S_{SU}$ is an effective method to discriminate folded proteins from natively unfolded ones in genomes. We have already attempted at establishing a scaling law of the number of the unfolded proteins with the size of a genome, getting a first estimate for the scaling exponent of 1.95 ± 0.21 [20]. We get back here to this problem taking into account a greater number of eukaryotic genomes.

We used the strictly unanimous score to evaluate the percentage of natively unfolded proteins in the genomes of several organisms selected from Archaea, Bacteria and Eukarya. The results are reported in the table given in the appendix. $S_{SU}$ predicts about 0.8% of archaean, 3.7% of eubacterial and 23.4% of eukaryotic proteins as natively unfolded. These results are consistent with those previously reported in the literature [21,22]. We note also that the percentage of unclassified proteins in the genomes is below 10% in Archaea and Bacteria, whereas is comprised between 10 and 20% in Eukarya.

In figure 1 we show the correlation among the number of natively unfolded proteins and the number of proteins in the genomes. We find a correlation coefficient of 0.94, that increases to 0.97 if we exclude Archaea. It is evident that, with the exception of *Halobacterium sp.*, archaean genomes tend to have less natively unfolded proteins than bacterial genomes of the same size. The scaling exponent relating the number of natively unfolded proteins to the number of proteins in the genomes is $1.81 \pm 0.10$.

A critical assessment about the reliability of these figures should be done, at this point. As explained above we have introduced the consensus score $S_{SU}$ to avoid, at the expenses of excluding a few proteins from the classification, overestimation of disordered proteins in a genome and conflict among single scoring indexes previously introduced. To quantitatively illustrate this point we present in table 2 the scaling exponents that we have obtained by evaluating, on the same set of genomes, the number of disordered proteins $N_d$ through the use of $S_{SU}$, mean packing, mean contact energy and gVSL2.

**Table 2 Scaling exponents for different disorder predictors**

| Scoring index | Scaling exponent | Correlation coefficient |
| --- | --- | --- |
| $S_{SU}$ | 1.81±0.10 | 0.97 |
| Mean packing | 1.59±0.07 | 0.98 |
| Mean contact energy | 1.66±0.08 | 0.97 |
| gVSL2 | 1.58±0.07 | 0.97 |

It is evident that the single indexes give values of the scaling exponent that coincide within the uncertainties, but differ from that determined through $S_{SU}$. The main difference between $S_{SU}$ and the other indexes resides that the former excludes some proteins from the classification, exactly those on which the other indexes would take conflicting decisions. It is then reasonable to attribute the systematic discrepancy between $S_{SU}$ and the other indexes to the noise or ambiguity due to the presence of the proteins that are not classified by $S_{SU}$. Following this line of reasoning we checked that re-evaluating the scaling exponents on the same set of genomes, but purged from the proteins filtered out by $S_{SU}$, gave scaling exponent that coincide with that estimated by $S_{SU}$. The set of proteins over which mean packing, mean contact energy and gVSL did not reach a consensus has an interest *per se*, as a set of structurally ambiguous proteins. We are ready to send the list upon request to interested readers. It is worth mentioning another interesting scaling relation between the number of proteins left unclassified by $S_{SU}$ and the total number of proteins in a genome: the scaling exponents is, in this case $1.29 \pm 0.05$, with a correlation coefficient of 0.99. and always excluding sequences from Archaea.

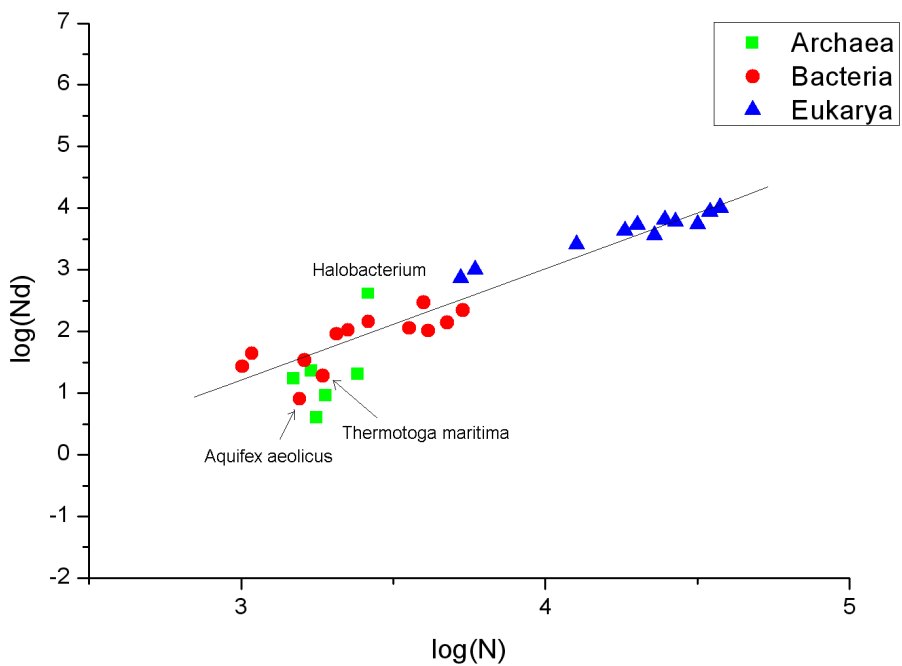

**Figure 1 - Number of predicted natively unfolded proteins vs. total number of proteins in various genomes**

Logarithmic plot of the number of natively unfolded proteins predicted by $S_{SU}$ vs. the total number of proteins in the genomes. The exponent of the power law 1.81 ± 0.10.

## Discussion and final remarks

The main point we address in this letter is that using single disorder predictors such as mean packing, mean contact energy and gVSL2 a bias in the estimated number of disordered proteins in a genome can be introduced. This bias can be cured through the introduction of the combination consensus index $S_{SU}$, which leaves a certain fraction of proteins unclassified. The difference in scaling exponents is removed by removing the proteins that escape the consensus. This observation is definitely a consistency check that gives a certain degree of reliability to our estimate of the scaling exponent of the number of disordered proteins with the size of the genomes.

Reverting to the biological meaning of our result we note that it has been previously reported that the percentage of natively unfolded proteins is higher in eukaryotic organisms with respect to Archaea and Bacteria [21]. This tendency has been related to the fact that eukarya have more complex regulatory and signalling networks, in which the presence of flexible proteins may be advantageous due to their ability to bind several targets with high specificity and low affinity. This idea is supported by the observation that several natively unfolded proteins are involved in regulatory post-transcriptional and post-translational processes [1,7-9]; moreover it has been reported that, in protein interactions networks, disorder is frequent in hub proteins [23-25].

At present we do not have an interpretation for the value of the scaling exponent we have found. We have shown that 1.81±0.10 is a robust estimate, but we cannot explain it. In the next future we shall extend our investigation by considering more genomes. It will be interesting to check if the scaling exponent and the percentage of unclassified proteins remain stable under enlargement of the dataset.

It is important to develop evolutionary models to understand the results in figure 1. To this end, it is important to note that Archaea seem to be exceptions to the scaling law connecting the number of natively unfolded proteins to the total number of proteins in a genome. Interestingly, most of the Archaea here analysed were thermophiles; moreover, two of the Bacteria that exhibited a low number of natively unfolded proteins were also thermophiles (*Aquifex Aeolicus* and *Thermotoga Maritima*), these observations support the idea that thermophilic organisms tend to adopt more rigid protein structures to afford high temperature environments [26]. It has been also suggested that Archaea separated early from the last common ancestor of all organisms, if this is true then they should have undergone a specific selective pressure to thrive in extreme environments [27]; that could explain why they do not follow the possibly universal scaling suggested by our results.

## Acknowledgments

The authors thank Prof. A. Colosimo for interesting comments during the preparation of the manuscript that helped us in clarifying some points.

# Appendix

**Table - Frequency of natively unfolded proteins in various genomes[1]**

| ORGANISM | N. proteins | $S_{SU}$ | |
|---|---|---|---|
| | | % predicted | % unclassified |
| **ARCHAEA** | | | |
| *Aeropyrum pernix* | 1700 | 1.3 | 5.3 |
| *Archaeoglobus fulgidus* | 2418 | 0.8 | 5.0 |
| *Halobacterium sp.*[2] | 2622 | 16.2 | 30.8 |
| *Methanococcus jannaschii* | 1768 | 0.2 | 5.4 |
| *Pyrococcus abyssii* | 1898 | 0.5 | 5.1 |
| *Thermoplasma volcanium* | 1491 | 1.1 | 4.5 |
| | **9275** | **0.8** | **5.1** |
| **BACTERIA** | | | |
| *Agrobacterium tumefaciens C58* | 5355 | 4.1 | 8.0 |
| *Aquifex aeolicus* | 1558 | 0.5 | 5.9 |
| *Clamydophila pneumoniae AR39* | 1085 | 4.1 | 9.0 |
| *Chlorobium tepidum TLS* | 2247 | 4.7 | 7.7 |
| *Escherichia coli K12* | 4130 | 2.5 | 6.1 |
| *Haemophilus influenzae Rd* | 1615 | 2.1 | 5.2 |
| *Mycobacterium tuberculosis H37Rv* | 3989 | 7.4 | 11.6 |
| *Neisseria meningitidis MC58* | 2063 | 4.4 | 8.3 |
| *Salmonella typhi* | 4756 | 3.0 | 6.6 |
| *Staphylococcus aureus COL* | 2618 | 5.5 | 6.9 |
| *Synechocystis species PCC 6803* | 3569 | 3.2 | 6.4 |
| *Thermotoga maritima* | 1856 | 1.0 | 5.8 |
| Treponema pallidum | 1009 | 2.7 | 6.7 |
| | **35850** | **3.7** | **7.4** |
| **EUKARYA** | | | |
| *Anopheles gambiae* | 12649 | 20.5 | 12.2 |
| *Arabidopsis thaliana* | 31708 | 17.5 | 14.6 |
| *Bos taurus* | 24686 | 26.3 | 15.4 |
| *Caernorhabditis elegans* | 22843 | 16.1 | 13.0 |
| *Drosophila melanogaster* | 20046 | 26.6 | 14.4 |
| *Homo sapiens* | 37412 | 27.5 | 18.6 |
| *Macaca mulatta* | 37606 | 26.9 | 16.4 |
| *Mus musculus* | 34699 | 25.1 | 16.9 |
| *Oryza sativa* | 26763 | 22.6 | 15.4 |
| *Plasmodium falciparum* | 5260 | 14.0 | 23.1 |
| *Saccharomyces cerevisiae* | 5880 | 17.0 | 14.2 |
| *Gallus gallus* | 18244 | 23.9 | 15.6 |
| | **277796** | **23.4** | **15.8** |

[1]genomes were download from the ftp server of NCBI: ftp://ftp.ncbi.nlm.nih.gov/genomes/
[2]Halobacterium sp. is an outlier, so we did not consider it in the computation of the mean of disordered proteins and unclassified proteins in the Archaea